\documentclass{emulateapj}

\shorttitle{Near-infrared thermal emission from TrES-2b}
\shortauthors{Croll et al.}

\newcommand{\FpOverFStarPercentAbstractTrESTwo}{0.062}
\newcommand{\FpOverFStarPercentAbstractMinusTrESTwo}{0.011}
\newcommand{\FpOverFStarPercentAbstractPlusTrESTwo}{0.013}
\newcommand{\XSigmaTrESTwo}{5}

\newcommand{\PhaseAbstractTrESTwo}{0.5014}
\newcommand{\PhaseAbstractMinusTrESTwo}{0.0013}
\newcommand{\PhaseAbstractPlusTrESTwo}{0.0013}
\newcommand{\ECosOmegaTrESTwo}{0.0020}
\newcommand{\ECosOmegaPlusTrESTwo}{0.0021}
\newcommand{\ECosOmegaMinusTrESTwo}{0.0021}
\newcommand{\ECosOmegaAbsoluteThreeSigmaLimitTrESTwo}{0.0090}
\newcommand{\TBrightTrESTwo}{1636}
\newcommand{\TBrightPlusTrESTwo}{79}
\newcommand{\TBrightMinusTrESTwo}{88}

\newcommand{\fReradiationTrESTwo}{0.358}
\newcommand{\fReradiationPlusTrESTwo}{0.074}
\newcommand{\fReradiationMinusTrESTwo}{0.072}

\newcommand{\cOneTrESTwo}{0.00061}
\newcommand{\cOnePlusTrESTwo}{0.00010}
\newcommand{\cOneMinusTrESTwo}{0.00012}
\newcommand{\cTwoTrESTwo}{-0.005}
\newcommand{\cTwoPlusTrESTwo}{0.001}
\newcommand{\cTwoMinusTrESTwo}{0.001}

\newcommand{\TOffsetTrESTwo}{4.5}

\newcommand{\TOffsetPlusTrESTwo}{4.8}
\newcommand{\TOffsetMinusTrESTwo}{4.6}

\newcommand{\ChiTrESTwo}{1.089}
\newcommand{\ChiPlusTrESTwo}{0.003}
\newcommand{\ChiMinusTrESTwo}{0.002}
\newcommand{\JDOffsetONETrESTwo}{14994.0605}
\newcommand{\JDOffsetPlusONETrESTwo}{0.0033}
\newcommand{\JDOffsetMinusONETrESTwo}{0.0032}

\newcommand{\FpOverFStarPercentAbstractTrESTwoCyclicZero}{0.064}
\newcommand{\FpOverFStarPercentAbstractMinusTrESTwoCyclicZero}{0.005}
\newcommand{\FpOverFStarPercentAbstractPlusTrESTwoCyclicZero}{0.005}

\newcommand{\PhaseAbstractTrESTwoCyclicZero}{0.5012}
\newcommand{\PhaseAbstractMinusTrESTwoCyclicZero}{0.0008}
\newcommand{\PhaseAbstractPlusTrESTwoCyclicZero}{0.0013}
\newcommand{\ECosOmegaTrESTwoCyclicZero}{0.0017}
\newcommand{\ECosOmegaPlusTrESTwoCyclicZero}{0.0020}
\newcommand{\ECosOmegaMinusTrESTwoCyclicZero}{0.0020}

\newcommand{\TBrightTrESTwoCyclicZero}{1646}
\newcommand{\TBrightPlusTrESTwoCyclicZero}{29}
\newcommand{\TBrightMinusTrESTwoCyclicZero}{30}

\newcommand{\fReradiationTrESTwoCyclicZero}{0.367}
\newcommand{\fReradiationPlusTrESTwoCyclicZero}{0.027}
\newcommand{\fReradiationMinusTrESTwoCyclicZero}{0.026}

\newcommand{\cOneTrESTwoCyclicZero}{0.00061}
\newcommand{\cOnePlusTrESTwoCyclicZero}{0.00008}
\newcommand{\cOneMinusTrESTwoCyclicZero}{0.00007}
\newcommand{\cTwoTrESTwoCyclicZero}{-0.005}
\newcommand{\cTwoPlusTrESTwoCyclicZero}{0.001}
\newcommand{\cTwoMinusTrESTwoCyclicZero}{0.002}

\newcommand{\TOffsetTrESTwoCyclicZero}{3.8}

\newcommand{\TOffsetPlusTrESTwoCyclicZero}{4.6}
\newcommand{\TOffsetMinusTrESTwoCyclicZero}{3.0}

\newcommand{\ChiTrESTwoCyclicZero}{1.086}
\newcommand{\ChiPlusTrESTwoCyclicZero}{0.018}
\newcommand{\ChiMinusTrESTwoCyclicZero}{0.001}
\newcommand{\JDOffsetONETrESTwoCyclicZero}{14994.0600}
\newcommand{\JDOffsetPlusONETrESTwoCyclicZero}{0.0032}
\newcommand{\JDOffsetMinusONETrESTwoCyclicZero}{0.0021}

\newcommand{\fReradiationTrESTwoALL}{0.346}
\newcommand{\fReradiationPlusTrESTwoALL}{0.038}
\newcommand{\fReradiationMinusTrESTwoALL}{0.037}
\newcommand{\BlackbodyOneChi}{9.1}	
\newcommand{\BlackbodyTwoChi}{4.7}	
\newcommand{\FortneyOneChi}{15.3}	
\newcommand{\FortneyTwoChi}{25.4}	
\newcommand{\FortneyThreeChi}{5.5}	
\newcommand{\FortneyFourChi}{10.6}	
\newcommand{\Lday}{5.3}
\newcommand{\DaysidePercentage}{69}
\newcommand{\DurbinWatsonTestStatistic}{1.97}
\newcommand{\XPointXX}{13.7$\times$10$^{-3}$}
\newcommand{\YPointYY}{0.71$\times$10$^{-3}$}
\newcommand{\BinNumberImages}{5}
\newcommand{\HowMany}{11}

\begin{document}

\title{Near-Infrared Thermal Emission from the Hot Jupiter TrES-2b:
Ground-Based Detection of the Secondary Eclipse\altaffilmark{*}}

\author{Bryce Croll\altaffilmark{1},
Loic Albert\altaffilmark{2},
David Lafreniere\altaffilmark{3},
Ray Jayawardhana\altaffilmark{1},
Jonathan J. Fortney\altaffilmark{4} 
}

\altaffiltext{1}{Deptartment of Astronomy and Astrophysics, University of Toronto, 50 St. George Street, Toronto, ON 
M5S 3H4, Canada;
croll@astro.utoronto.ca}

\altaffiltext{2}{Canada-France-Hawaii Telescope Corporation, 65-1238 Mamalahoa Highway,
Kamuela, HI 96743.}

\altaffiltext{3}{D\'epartement de physique, Universit\'e de Montr\'eal, C.P.
6128 Succ. Centre-Ville, Montr\'eal, QC, H3C 3J7, Canada}

\altaffiltext{4}{Department of Astronomy and Astrophysics, University of California, Santa Cruz, CA, 95064}


\altaffiltext{*}{Based on observations obtained with WIRCam, a joint project of CFHT, Taiwan, Korea, Canada, France, at the Canada-France-Hawaii Telescope (CFHT) which is operated by the National Research Council (NRC) of Canada, the Institute National des Sciences de l'Univers of the Centre National de la Recherche Scientifique of France, and the University of Hawaii.}

\begin{abstract}

We present near-infrared Ks-band photometry bracketing the secondary eclipse of the hot Jupiter TrES-2b
using the Wide-field Infrared Camera on the Canada-France-Hawaii Telescope.
We detect its thermal emission with an eclipse depth of 
\FpOverFStarPercentAbstractTrESTwo$^{+\FpOverFStarPercentAbstractPlusTrESTwo}_{-\FpOverFStarPercentAbstractMinusTrESTwo}$\%
(\XSigmaTrESTwo $\sigma$). 
Our best-fit secondary eclipse is consistent with a circular orbit
(a 3$\sigma$ upper limit on the eccentricity, $e$, and argument or periastron, $\omega$, of 
$|$$e$$\cos$$\omega$$|$ $<$ \ECosOmegaAbsoluteThreeSigmaLimitTrESTwo),
in agreement with mid-infrared detections of the secondary eclipse of this planet.
A secondary eclipse of this depth corresponds to a day-side Ks-band
brightness temperature of $T_B$ = \TBrightTrESTwo$^{+\TBrightPlusTrESTwo}_{-\TBrightMinusTrESTwo}$ $K$.
Our thermal emission measurement when combined with the thermal emission measurements 
using Spitzer/IRAC from O'Donovan and collaborators
suggest that this planet exhibits relatively efficient
day to night-side redistribution of heat
and a near isothermal dayside atmospheric temperature structure, with a spectrum
that is well approximated by a blackbody. 
It is unclear if the atmosphere of TrES-2b
requires a temperature inversion; if it does it is likely due to 
chemical species other than TiO/VO as the atmosphere of TrES-2b is too cool
to allow TiO/VO to remain in gaseous form.
Our secondary eclipse has the smallest depth of any detected from the ground at around 2 $\mu m$ to date.

%


\end{abstract}


\section{Introduction}

The first detection of the transit of an exoplanet in front of its parent
star (\citealt{Charb00}; \citealt{Henry00}) opened a new avenue to 
determine the characteristics of these exotic worlds.
For all but the most eccentric cases, approximately
half-an-orbit after their transits
these planets pass behind their star along our line of sight
allowing their thermal flux to be measured in the infrared.
The first detections of an exoplanet's thermal emission 
(\citealt{Charb05}; \citealt{Deming05}) came from observations in space
with Spitzer using the Infrared Array Camera (IRAC; \citealt{Fazio04}). 
Since then the vast majority of 
such measurements 
have been made using Spitzer at wavelengths 
longer than 3 $\mu m$, and thus longwards of the blackbody peak of these 
``hot'' exoplanets. Recent observations have extended secondary eclipse detections 
into the near-infrared; the first detection was from space with NICMOS on 
the Hubble Space Telescope (\citealt{Swain09} at $\sim$2 $\mu m$). 
More recently, near-infrared 
detections have been achieved from the ground; the first of these detections
include
a $\sim$6$\sigma$ detection in K-band of TrES-3b 
using the William Herschel Telescope \citep{deMooij09},
a $\sim$4$\sigma$ detection in z'-band
of OGLE-TR-56b using Magellan and the Very Large Telescope (VLT; \citealt{Sing09}),
and a $\sim$5$\sigma$ detection at $\sim$2.1 $\mu m$ of
CoRoT-1b also with the VLT \citep{Gillon09}. 

Thermal emission measurements in the near-infrared are crucial to our understanding of these planets'
atmospheres, as they allow us to constrain hot Jupiters' thermal emission near their blackbody
peaks. The combination of Spitzer/IRAC and near-infrared thermal emission measurements allows us to constrain
the temperature-pressure profiles of these planets' atmospheres over a range of pressures \citep{Fortney08},
better estimate the bolometric luminosity of these planets' dayside emission,
and thus contributes to a more complete understanding of how these planets transport heat from the day to nightside at a variety
of depths and pressures in their atmospheres \citep{Barman08}.

 The transiting hot Jupiter TrES-2b orbits a G0 V star with a 
period of $\sim$2.47 $d$ \citep{ODonovan06}.
According to the \citet{Fortney08} theory this places TrES-2b marginally in the 
hottest, mostly highly irradiated
class (the pM-class) of hot Jupiters and
close to the dividing line between this hottest class and 
the merely warm class of hot Jupiters (the pL-class).
Thus TrES-2b could be a key object to refine the dividing line between
these two classes, 
and indicate
the physical cause of this demarcation, or reveal
whether this divide even exists.
Recently \citet{ODonovan09} used Spitzer/IRAC to measure the depth 
of the secondary eclipse of TrES-2b in the four IRAC bands.
Their best-fit eclipses are consistent with a circular orbit, and collectively they
are able to place a 3$\sigma$ limit
on the eccentricity, $e$, and argument of periastron, $\omega$, of $|$$e$cos$\omega$$|$ $<$ 0.0036.
Their best-fit eclipses at 3.6, 5.8 and 8.0 $\mu m$ are well-fit by a blackbody.
At 4.5 $\mu m$ they detect excess emission, in agreement with the
theory of several researchers (\citealt{Fortney08, Burrows08}) that predicts 
such excess due to water emission, rather than absorption,
at this wavelength due to a temperature inversion
in the atmosphere.
One-dimensional radiative-equilibrium models for hot Jupiter planets
generally show that the atmospheric opacity is dominated by water vapor,
which is especially high in the mid-infrared, but has prominent windows (the
JHK bands) in the near infrared \citep{Fortney08,Burrows08}. One can probe more deeply, to gas at
higher pressure, in these opacity windows.  Models without temperature
inversions feature strong emission in the JHK bands, since one sees down to
the hotter gas.  Models with temperature inversions, since they feature a
relatively hotter upper atmosphere and relatively cooler lower atmosphere,
yield weaker emission in the near-IR (JHK), but stronger emission in the
mid-infrared \citep{Hubeny03,Fortney06}.
Near-infrared thermal emission measurements should thus be useful to determine whether TrES-2b
does or does not harbour a temperature inversion.

Owing to its high irradiation, with an incident flux of $\sim$$1.1$$\times$$10^{9}$ $erg$$s^{-1}$$cm^{-2}$,
and favourable planet-to-star radius ratio ($R_{P}/R_{*}$$\sim$0.13),
we included TrES-2b in our program observing the secondary eclipses
of some of the hottest of the hot Jupiters from the ground. 
Here we present
Ks-band observations bracketing TrES-2b's secondary eclipse using
the Wide-field InfraRed Camera (WIRCam) on the Canada-France-Hawaii Telescope (CFHT).
We report a \XSigmaTrESTwo $\sigma$ detection of its thermal emission.

\section{Observations and data reduction}
\label{SecReduction}

 We observed TrES-2 ($K$=9.846) with WIRCam \citep{Puget04} on CFHT on 2009 June 10
under photometric conditions.
The observations lasted for $\sim$3.5 hours
evenly bracketing the predicted secondary eclipse of this hot Jupiter assuming it has a circular orbit.
Numerous reference stars were also observed in the 21x21 arcmin field of view of WIRCam.
To minimize the impact of flat field errors, intrapixel variations
and to keep the flux of the target star well
below detector saturation, we defocused the telescope to 1.5mm, such that 
the flux of our target was spread
over a ring $\sim$20 pixels in diameter (6\arcsec) on our array.

We observed TrES-2 in ``stare'' mode on CFHT where the target star is observed continuously without dithering.
5-second exposures were used to avoid saturation. To increase the observing efficiency
we acquired a series of data-cubes each containing twelve 5-second exposures. 
The twelve exposure data-cube is the maximum number of exposures allowed in a guide-cube
in queue mode at CFHT.
To counteract drifts in the position of the stars positions on the WIRCam chips,
which we had noticed in earlier WIRCam observations of secondary eclipses \citep{CrollTrESThree}, 
we initiated a corrective guiding ``bump'' before every image
cube to recenter the stellar point-spread-function as near as possible to the original pixels at the start of the observation.
The effective duty cycle
after accounting for readout and for saving exposures was 43\%.
The images were preprocessed with the `I`iwi
pipeline\footnote{\tiny http://www.cfht.hawaii.edu/Instruments/Imaging/WIRCam/IiwiVersion1Doc.html}. \normalsize
This pipeline includes the following steps: applying a non-linearity
flux correction, removing bad and saturated pixels, dark
subtraction, flat-fielding, sky subtraction, zero-point calibration and a
rough astrometry determination.
We sky subtract our data by constructing a normalized sky frame built by taking the median 
of a stack of source-masked and background-normalized on-sky images. 
Our on-sky images consist of 15 dithered in-focus images observed before
and after the on-target sequence.
For each on-target image the normalized sky frame is scaled to
the target median background level and then subtracted.

\begin{figure*}
\centering
\includegraphics[scale=0.50,angle=90]{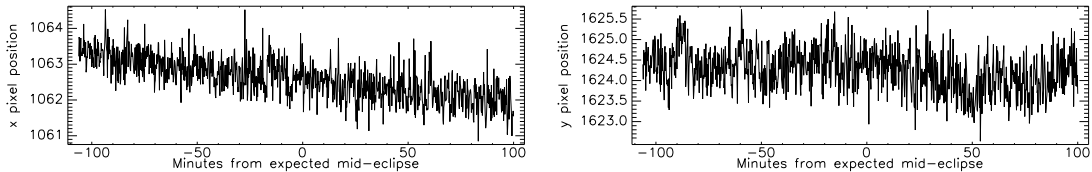}
\caption{	The x and y position of the centroid of the PSF of the target star, TrES-2, with time (top panels).}
\label{FigShifts}
\end{figure*}

 We performed aperture photometry on our target star and all unsaturated, reasonably bright reference stars on the WIRCam array.
We used a circular aperture with a radius of 12.5 pixels. 
We tested larger and smaller apertures in increments of 0.5 pixels, and confirmed that this size
of aperture returned optimal photometry.
The residual background was estimated using an annulus with an inner radius of 21 pixels, and an outer radius of 30 pixels; a few different sizes
of sky annuli were tested, and it was found that the accuracy of the resulting photometry
was not particularly sensitive to the size of the sky aperture.
As TrES-2 has a nearby reference star (0.17\arcmin \ separation) that falls in our sky aperture, we exclude a slice of the annulus
that falls near this reference star
to avoid any bias in background determination\footnote{Pixels that fall from 5$^{o}$ to -45$^{o}$ degrees as measured from due North towards the East are excluded from our annulus.}.
During our observations, despite the aforementioned corrective ``bump'' to keep the centroid of our
stellar point-spread-function (PSF) as steady as possible, our target star and the rest of
the stars on our array displayed high frequency
shifts in position (Figure \ref{FigShifts}).
To ensure that the apertures for our photometry were centered in the middle of the stellar PSFs,
we used a center-of-mass calculation, with pixel flux substituted for mass,
to determine the x and y center of our defocused stellar rings 
for each one of our target and reference star apertures.

\begin{figure*}
\centering
\includegraphics[scale=0.80,angle=90]{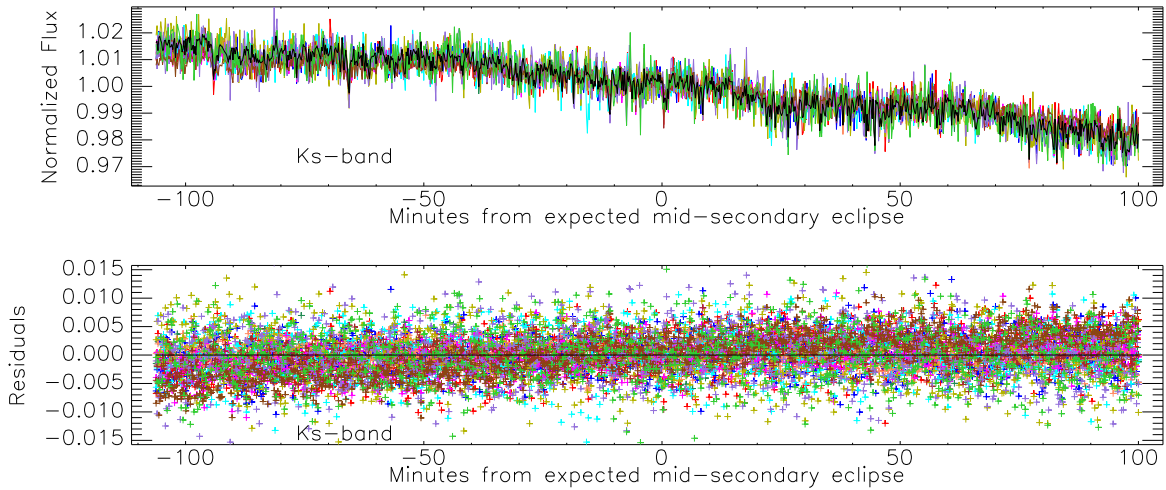}
\caption{	Top panel: The normalized flux from the target star (black) and the reference stars
		that are used to calibrate the flux of TrES-2b (various colours).
		Bottom panel: The residuals from the normalized flux of the target star of the
		normalized flux of the reference stars (various colours).
		The residuals have not been corrected for the x/y pixel positions of the target.}
\label{FigTrES2bRefStars}
\end{figure*}

The light curves for our target and reference stars following our
aperture photometry displayed significant, systematic variations in intensity 
(see the top panel of Figure \ref{FigTrES2bRefStars}), possibly due to changes in
atmospheric transmission, seeing and airmass, guiding errors and/or other effects.
The target light curve
was then corrected for these systematic variations by normalizing its flux to the 
\HowMany \ 
reference stars that show the smallest deviation from the target star outside of the expected occultation.
Reference stars that showed significant deviations in-eclipse
from that of the target star and other
reference stars, as indicated by a much larger root-mean-square in-eclipse than out-of-eclipse 
due to intermittent systematic effects for instance, were also excluded.
For the reference stars that were chosen for the comparison to our target star, the flux of each
one of these 
star was divided by its median-value, and then an average 
reference star light curve was produced by taking the mean of the lightcurves of these median-corrected
reference stars. The target
flux was then normalized by this mean reference star light curve.

\begin{deluxetable}{ccc}
\tabletypesize{\footnotesize}
\tablecaption{Reference Stars}
\tablehead{
\colhead{Reference Star \#}	&	\colhead{2MASS Identifier} & \colhead{K-band Magnitude} \\
}
\startdata
1 	&	 J19072977+4918354 	&	 10.294 	\\
2 	&	 J19065501+4916195 	&	 10.737 	\\
3 	&	 J19071365+4912041 	&	 11.270 	\\
4 	&	 J19065809+4916315 	&	 9.875 	\\
5 	&	 J19070093+4917323 	&	 11.337 	\\
6 	&	 J19074435+4915418 	&	 10.766 	\\
7 	&	 J19071824+4916526 	&	 11.239 	\\
8 	&	 J19073380+4916035 	&	 10.712 	\\
9 	&	 J19071955+4911176 	&	 11.514 	\\
10 	&	 J19075629+4923281 	&	 9.671 	\\
11 	&	 J19065548+4925404 	&	 11.454 	\\
\enddata
\label{TableStars}
\end{deluxetable}

\begin{figure}
\centering
\includegraphics[scale=0.43,angle=90]{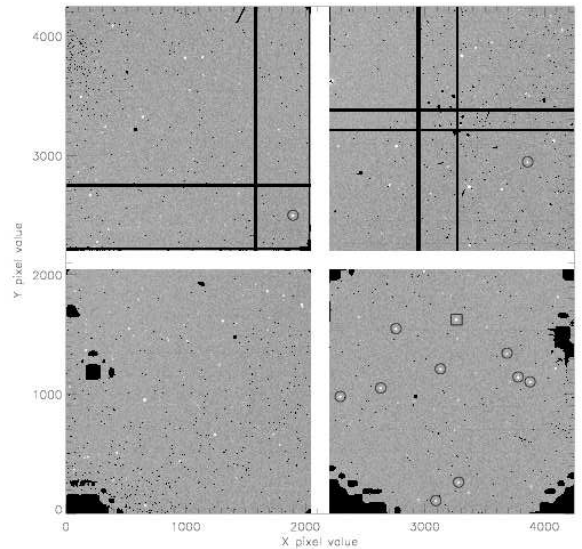}
\caption{	The CFHT/WIRCam full frame array during our observations of TrES-2b. The image has been preprocessed
		with the `I`iwi pipeline; the obvious artifacts (in the lower-left corners of the bottom two chips
		for instance or the crosses on the upper two chips) are due to the masking of bad pixels.
		The target star, TrES-2 (large square on the lower-right chip), and the reference stars 
		used to correct the flux of TrES-2 (circles) are marked.}
\label{FigTrES2bFullFrameArray}
\end{figure}

Figure \ref{FigTrES2bFullFrameArray} marks the \HowMany \ reference stars used to 
correct the flux of our target; the 2MASS identifiers of the reference stars are given in Table \ref{TableStars}.
Note that the majority of the reference stars with the smallest out of occultation
residuals to our target star are on the same chip as our target, despite the fact 
that there are other reference
stars on other chips closer in magnitude to our target. We believe this is due to 
the differential electronic 
response of the different WIRCam chips, and have noticed this same effect 
with other WIRCam observations
of other hot Jupiter secondary eclipses \citep{CrollTrESThree}.

\begin{figure*}
\centering
\includegraphics[scale=0.40,angle=90]{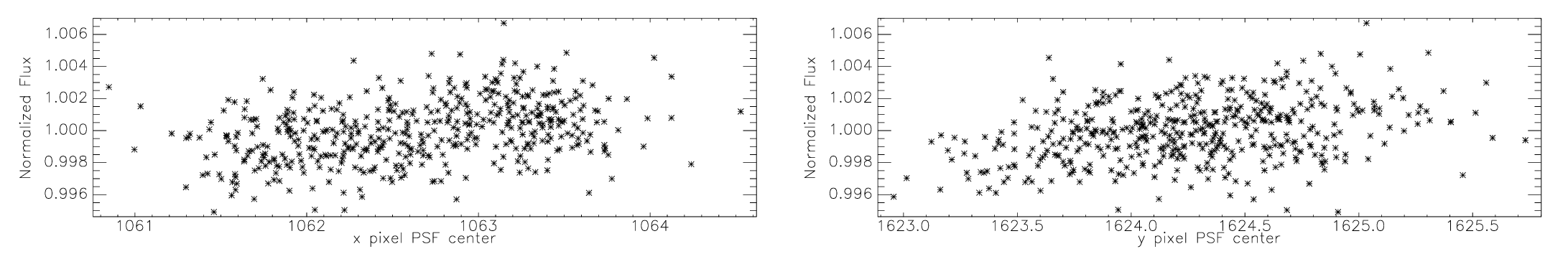}
\includegraphics[scale=0.40,angle=90]{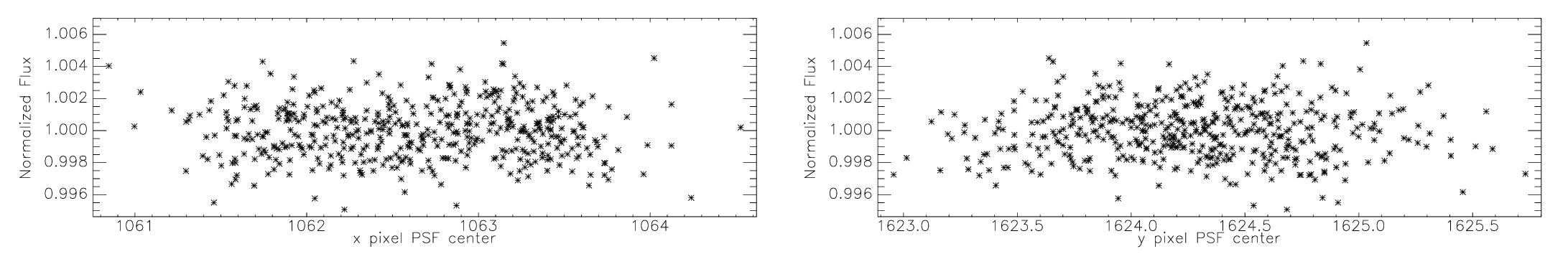}
\caption{	The out-of-eclipse photometry prior to the correction for the x and y position of the PSF (top panels).
		The out-of-eclipse photometry following this correction (bottom panels; see $\S$\ref{SecReduction} for details).}
\label{FigShiftsCorrection}
\end{figure*}

Following this correction we noticed that the flux of our target and reference stars 
displayed near-linear correlations with the x or y position of the centroid of the stellar PSF on the chip.
Given the aforementioned high frequecy of these shifts (Fig. \ref{FigShifts})
this suggests
that any leftover trend with position and the flux of the star was instrumental in origin.
Thus these near-linear trends (Figure \ref{FigShiftsCorrection} top panels) were removed
from the data for both the target and reference stars by performing a 
fit to the x and y position of the centroid of the PSF and the normalized flux
for the out-of-eclipse photometry.
We fit the out-of-eclipse photometric flux to the x and y position of the centroid of the PSF
with a function of the following form:
\begin{equation}
f = 1 + d_1 x + d_2 y + d_3 xy
\end{equation}
where $d_1$, $d_2$, and $d_3$ are constants.
We then apply this correction to both the in-eclipse and out-of-eclipse photometry.
The out-of-eclipse photometric data
prior to and following this correction are displayed in Figure \ref{FigShiftsCorrection} (bottom panels).
No other trends that were correlated with instrumental parameters were found.

\begin{figure}
\centering
\includegraphics[scale=0.45, angle = 270]{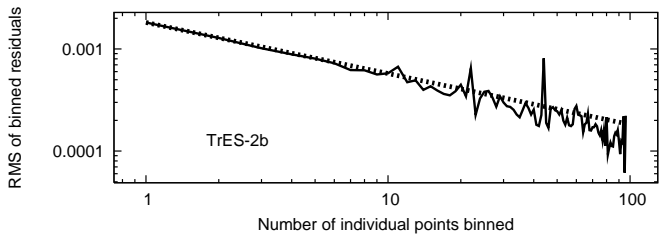}
\caption{	The root-mean-square of our out-of-eclipse photometry (solid line) 
		following the various corrections documented in $\S$\ref{SecReduction}.
		The dashed line displays the one over the square-root of the bin-size expectation for gaussian noise.
		}
\label{FigBinByN}
\end{figure}

By correcting the flux of our target with these \HowMany \
reference stars
and by removing the above correlation with the x/y position on the chip
the point-to-point scatter of our data outside occultation
improves from a root-mean-square (RMS)
of \XPointXX \ to \YPointYY \ per every 58 $s$ (or \BinNumberImages \ images).
The photometry following the aforementioned
analysis is largely free of systematics, as evidenced by the fact that the
out-of-eclipse photometric precision lies near the gaussian noise expectation for binning the data
of one over the square-root of the bin-size (Figure \ref{FigBinByN}).
Our observations in Ks-band, though,
are still well above the predicted photon noise RMS limit of 2.3$\times$10$^{-4}$ per 58 seconds. 
For the following analysis we set the uncertainty on our individual measurements as 0.95 times 
the RMS of the out of eclipse photometry after the removal of a linear-trend with time; we found
simply using 1.0 times the RMS of the out-of-eclipse photometry resulted in a reduced $\chi^{2}$ below one, and
thus resulted in a slight over-estimate of our errors.

\section{Analysis}

\begin{figure}
\centering
\includegraphics[scale=0.65,angle=270]{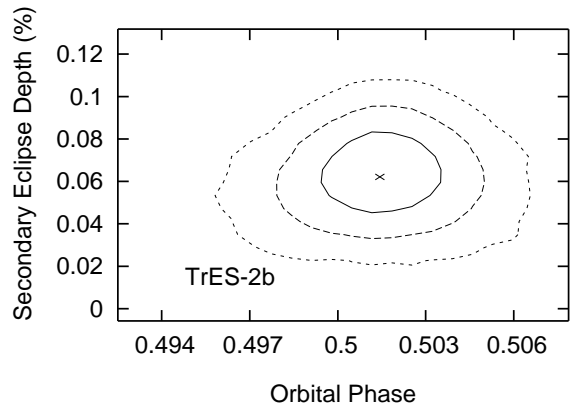}
\caption{	The 68.3\% (1$\sigma$; solid-line), 95.5\% (2$\sigma$; dashed-line)
		and 99.7\% (3$\sigma$; short dashed-line) credible regions
		from our MCMC analysis on the secondary eclipse depth, $\Delta F$,
		and phase, $\phi$. The ``x'' in the middle of the plot denotes the best-fit point from our 
		MCMC analysis.}
\label{FigTrES2bContour}
\end{figure}

\begin{figure}
\centering
\includegraphics[scale=0.35,angle=270]{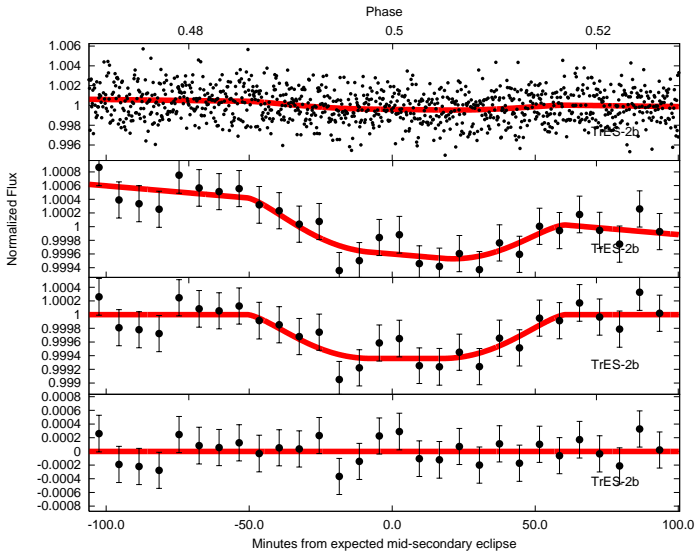}
\caption{	CFHT/WIRCam Ks-band photometry of the secondary eclipse of TrES-2b.
		The top panel shows the unbinned lightcurve, while the panel that is the second from the top shows the lightcurve
		with the data binned every 7.0 minutes.
		The panel that is the second
		from the bottom shows the binned data after the subtraction of
		the best-fit background, $B_{f}$, while the bottom 
		panel shows the binned residuals from the best-fit model.
		In each one of the panels the best-fit best-fit secondary eclipse and background, $B_f$,
		is shown with the red line.
		The expected mid-secondary eclipse is if TrES-2b has zero eccentricity.
	}
\label{FigTrES2b}
\end{figure}

Similarly to nearly all our near-infrared photometric data-sets taken with CFHT/WIRCam (e.g. \citealt{CrollTrESThree,CrollWASPTwelve}),
our Ks-band photometry following the reduction exhibited an obvious background trend, $B_f$. This background term displayed
a near-linear slope, and thus we fit the background with a linear-function of the form:
\begin{equation}
B_f = 1 + c_1 + c_2 dt
\end{equation}
where dt is the time interval from the beginning of the observations, and $c_1$ and $c_2$ are the fit parameters.
Given that most of our other data-sets display these background trends, it is unlikely, but not impossible, that this
slope is intrinsic to TrES-2.
We fit for the best-fit secondary eclipse and linear fit simultaneously 
using Markov Chain Monte Carlo (MCMC) methods
(\citealt{Christensen01}; \citealt{Ford}; described
for our purposes in \citealt{CrollMCMC}). We use a 5$\times$10$^{6}$ step MCMC chain. We fit
for $c_1$, $c_2$, the depth of the secondary eclipse, $\Delta F$, and the offset
that the eclipse occurs later than the 
expected eclipse center, $t_{\rm offset}$\footnote{we take into account the 0.6 minute offset due to light travel-time in the system [\citealt{Loeb05}]}.
We also quote the best-fit phase, $\phi$, 
as well as the best-fit mid-eclipse heliocentric julian-date, $t_{eclipse}$.
We use the \citet{Mandel02} algorithm without limb darkening to generate our best-fit secondary eclipse model. 
We obtain our stellar and planetary parameters for TrES-2 from \citet{Torres08}, including the planetary period and ephemeris.
The results from these fits are presented in Table \ref{TableParams}.
The phase dependence of the best-fit secondary eclipse
is presented in Figure \ref{FigTrES2bContour}.
The best-fit secondary eclipse is presented in Figure \ref{FigTrES2b}. 

\begin{deluxetable}{ccccc}
\tabletypesize{\footnotesize}
\tablecaption{Best-fit secondary eclipse parameters}
\tablehead{
\colhead{Parameter} 	& \colhead{MCMC} 	& \colhead{``Residual} 		\\	
\colhead{}		& \colhead{Solution}	& \colhead{ permutation''}	\\
\colhead{}		& \colhead{}		& \colhead{Solution}		\\
}
\startdata
reduced $\chi^{2}$			&	\ChiTrESTwo$^{+\ChiPlusTrESTwo}_{-\ChiMinusTrESTwo}$										& \ChiTrESTwoCyclicZero$^{+\ChiPlusTrESTwoCyclicZero}_{-\ChiMinusTrESTwoCyclicZero}$ 										\\
$\Delta F$				&	\FpOverFStarPercentAbstractTrESTwo$^{+\FpOverFStarPercentAbstractPlusTrESTwo}_{-\FpOverFStarPercentAbstractMinusTrESTwo}$\%	& \FpOverFStarPercentAbstractTrESTwoCyclicZero$^{+\FpOverFStarPercentAbstractPlusTrESTwoCyclicZero}_{-\FpOverFStarPercentAbstractMinusTrESTwoCyclicZero}$\%	\\
$t_{offset}$ ($min$)\tablenotemark{a}	&	\TOffsetTrESTwo$^{+\TOffsetPlusTrESTwo}_{-\TOffsetMinusTrESTwo}$								& \TOffsetTrESTwoCyclicZero$^{+\TOffsetPlusTrESTwoCyclicZero}_{-\TOffsetMinusTrESTwoCyclicZero}$ 								\\
$t_{eclipse}$ (HJD-2440000)		&	\JDOffsetONETrESTwo$^{+\JDOffsetPlusONETrESTwo}_{-\JDOffsetMinusONETrESTwo}$							& \JDOffsetONETrESTwoCyclicZero$^{+\JDOffsetPlusONETrESTwoCyclicZero}_{-\JDOffsetMinusONETrESTwoCyclicZero}$							\\
$c_1$					&	\cOneTrESTwo$^{+\cOnePlusTrESTwo}_{-\cOneMinusTrESTwo}$										& \cOneTrESTwoCyclicZero$^{+\cOnePlusTrESTwoCyclicZero}_{-\cOneMinusTrESTwoCyclicZero}$ 									\\
$c_2$ ($d^{-1}$)			&	\cTwoTrESTwo$^{+\cTwoPlusTrESTwo}_{-\cTwoMinusTrESTwo}$										& \cTwoTrESTwoCyclicZero$^{+\cTwoPlusTrESTwoCyclicZero}_{-\cTwoMinusTrESTwoCyclicZero}$ 									\\
$\phi$	\tablenotemark{a}		&	\PhaseAbstractTrESTwo$^{+\PhaseAbstractPlusTrESTwo}_{-\PhaseAbstractMinusTrESTwo}$						& \PhaseAbstractTrESTwoCyclicZero$^{+\PhaseAbstractPlusTrESTwoCyclicZero}_{-\PhaseAbstractMinusTrESTwoCyclicZero}$						\\
$T_{B}$	($K$)				&	\TBrightTrESTwo$^{+\TBrightPlusTrESTwo}_{-\TBrightMinusTrESTwo}$ 								& \TBrightTrESTwoCyclicZero$^{+\TBrightPlusTrESTwoCyclicZero}_{-\TBrightMinusTrESTwoCyclicZero}$  								\\
$e \cos(\omega)$ \tablenotemark{a}	&	\ECosOmegaTrESTwo$^{+\ECosOmegaPlusTrESTwo}_{-\ECosOmegaMinusTrESTwo}$								& \ECosOmegaTrESTwoCyclicZero$^{+\ECosOmegaPlusTrESTwoCyclicZero}_{-\ECosOmegaMinusTrESTwoCyclicZero}$ 								\\
$f_{Ks}$					& 	\fReradiationTrESTwo$^{+\fReradiationPlusTrESTwo}_{-\fReradiationMinusTrESTwo}$							& \fReradiationTrESTwoCyclicZero$^{+\fReradiationPlusTrESTwoCyclicZero}_{-\fReradiationMinusTrESTwoCyclicZero}$						\\
\enddata
\tablenotetext{a}{We account for the increased light travel-time in the system \citep{Loeb05}.}
\label{TableParams}
\end{deluxetable}


To determine the effect of any excess systematic noise on our photometry and the resulting fits
we employ the ``residual-permutation'' method
as discussed in \citet{Winn09}.
In this method the best-fit model is subtracted from the data, the residuals
are shifted between
1 and the total number of data points ($N$=1056 in our case),
and then the best-fit model is added back to the residuals. 
We then refit the adjusted lightcurve with a 5000-step MCMC chain
and record the parameters of the lowest $\chi^2$ point reached.
By inverting the residuals
we are able to perform $2 N-1$ total iterations.
The best-fit parameters and uncertainties obtained with this method
are similar to those found for our MCMC method
and are listed in Table \ref{TableParams}.
As the two methods produce similar results we
employ the MCMC errors for the rest of this paper.
We also test for autocorrelation among the residuals to our best-fit model using the Durbin-Watson 
test \citep{DurbinWatson51}; for the Durbin-Watson test 
a test-statistic greater than 1.0 and less than 3.0 (ideally near 2.0) 
indiciates a lack of autocorrelation. Our residuals pass this test with a test-statistic
of \DurbinWatsonTestStatistic.

\section{Discussion}

 The depth of our best-fit secondary eclipse is 
\FpOverFStarPercentAbstractTrESTwo$^{+\FpOverFStarPercentAbstractPlusTrESTwo}_{-\FpOverFStarPercentAbstractMinusTrESTwo}$\%.
The reduced $\chi^{2}$ is \ChiTrESTwo. Our best-fit secondary eclipse is consistent with a circular orbit;
the offset
from the expected eclipse center is: $t_{offset}$ = \TOffsetTrESTwo$^{+\TOffsetPlusTrESTwo}_{-\TOffsetMinusTrESTwo}$ minutes
(or at a phase of $\phi$=\PhaseAbstractTrESTwo$^{+\PhaseAbstractPlusTrESTwo}_{-\PhaseAbstractMinusTrESTwo}$).
This corresponds to a limit on the eccentricity and argument of periastron of
$e \cos \omega$ = \ECosOmegaTrESTwo$^{+\ECosOmegaPlusTrESTwo}_{-\ECosOmegaMinusTrESTwo}$, or a 3$\sigma$
limit of 
$|$$e$$\cos$$\omega$$|$ $<$ \ECosOmegaAbsoluteThreeSigmaLimitTrESTwo).
Our result is fully consistent with the more sensitive $e$cos$\omega$ limits
reported from the secondary eclipse detections at the four Spitzer/IRAC wavelengths \citep{ODonovan09}.
Thus our result bolsters the conclusion of \citet{ODonovan09} that tidal damping of the orbital eccentricity is unlikely to be responsible
for ``puffing up'' the radius of this exoplanet.

A secondary eclipse depth of 
\FpOverFStarPercentAbstractTrESTwo$^{+\FpOverFStarPercentAbstractPlusTrESTwo}_{-\FpOverFStarPercentAbstractMinusTrESTwo}$\%
corresponds to a brightness temperature of $T_{B}$ = \TBrightTrESTwo$^{+\TBrightPlusTrESTwo}_{-\TBrightMinusTrESTwo}$ $K$
in the Ks-band assuming the planet radiates as a blackbody, and adopting a stellar effective
temperature of $T_{eff}$ = 5850 $\pm$ 50 \citep{Sozzetti07}. 
This compares to the equilibrium temperature of TrES-2b 
of $T_{eq}$$\sim$1472 $K$ assuming isotropic reradiation, and a zero Bond albedo.
Hot Jupiter thermal emission measurements allow joint constraints on the Bond albedo, $A_B$,
and the efficiency of day to nightside redistribution of heat on these
presumably tidally locked planets.
The Bond albedo, $A_B$ is the fraction of the bolometric, incident stellar irradiation that is reflected by the planet's atmosphere.
We parameterize the redistribution of dayside stellar radiation absorbed by the planet's atmosphere to
the nightside by the reradiation factor,
$f$, following the \citet{Lopez07} definition.
If we assume a Bond albedo near zero, consistent with observations of other hot Jupiters \citep{Charbonneau99,Rowe08}
and with model predictions \citep{Burrows08b},
we find a reradiation factor of $f_{Ks}$ = \fReradiationTrESTwo$^{+\fReradiationPlusTrESTwo}_{-\fReradiationMinusTrESTwo}$
from our Ks-band eclipse photometry only,
indicative of relatively efficient advection of heat from the day-to-nightside at this wavelength.
In comparison, the reradiation factor for an atmosphere that reradiates isotropically is 
$f$=$\frac{1}{4}$, while $f$=$\frac{1}{2}$ denotes redistribution and reradiation over the dayside 
face only.

\begin{figure}
\centering
\includegraphics[scale=0.65,angle=270]{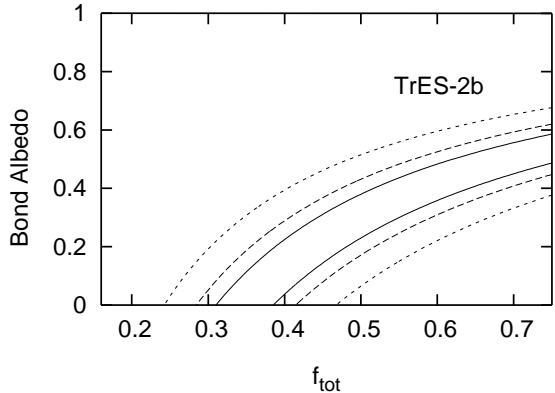}
\caption{	The 68.3\% (1$\sigma$; solid-line), 95.5\% (2$\sigma$; dashed-line)
		and 99.7\% (3$\sigma$; short dashed-line) $\chi^2$ confidence regions
		on the reradiation factor, $f_{tot}$, and Bond albedo from
		the combination of our Ks-band point and the
		Spitzer/IRAC measurements \citep{ODonovan09}.}
\label{FigBondReradiation}
\end{figure}

Our secondary eclipse depth,
when combined with the secondary eclipse depths at the Spitzer/IRAC wavelengths from
\citet{ODonovan09}, is consistent with a range of Bond albedos, $A_B$, and efficiencies of the day to nightside redistribution
on this presumably tidally locked planet (Figure \ref{FigBondReradiation}).
The best-fit total reradiation factor, $f_{tot}$, 
that results from a $\chi^{2}$ analysis of all the eclipse depths for TrES-2b
assuming a zero Bond albedo is $f_{tot}$ = \fReradiationTrESTwoALL$^{+\fReradiationPlusTrESTwoALL}_{-\fReradiationMinusTrESTwoALL}$. 
Thus our Ks-band brightness 
temperature ($T_{B}$ = \TBrightTrESTwo$^{+\TBrightPlusTrESTwo}_{-\TBrightMinusTrESTwo}$ $K$) and reradiation
factor $f_{Ks}$=\fReradiationTrESTwo$^{+\fReradiationPlusTrESTwo}_{-\fReradiationMinusTrESTwo}$,
reveal an atmospheric layer that is similar to, and perhaps slightly hotter, than the atmospheric layers 
probed by longer wavelength Spitzer observations
($T_B$$\sim$1500 $K$ from Spitzer/IRAC observations of TrES-2b [\citealt{ODonovan09}]).
The Ks-band is expected to be at a minimum in the water opacity \citep{Fortney08,Burrows08}, and thus our
Ks-band observations are expected to be able to see deep into the atmosphere of TrES-2b.
Our observations suggest that the deep, high pressure atmosphere of TrES-2b displays a similar temperature -- 
perhaps a slightly warmer temperature -- to lower pressure regions.

Another way of parameterizing the level of day-to-nightside heat redistribution is calculating the percentage 
of the bolometric luminosity emitted by the planet's dayside, $L_{day}$, compared to the 
nightside emission, $L_{night}$. 
Measurements of the thermal emission of a hot Jupiter at its blackbody peak
provide a valuable constraint on the bolometric luminosity of the planet's
dayside emission, and by inference its nightside emission \citep{Barman08}.
From simple thermal equilibrium arguments if
TrES-2b has a zero Bond albedo and it is in thermal equilibrium with its surroundings 
it should have a total bolometric luminosity of
$L_{tot}$ = 7.7$\times$10$^{-5}$$L_{\odot}$. 
By integrating the luminosity per unit frequency of our best-fit blackbody model across a wide wavelength range
we are able to calculate the percentage of the total luminosity reradiated by the dayside as $\sim$\DaysidePercentage\%
($L_{day}$ = \Lday$\times$10$^{-5}$$L_{\odot}$). The remainder, presumably, is advected via winds to the nightside.

\begin{figure}
\centering
\includegraphics[scale=0.465,angle=270]{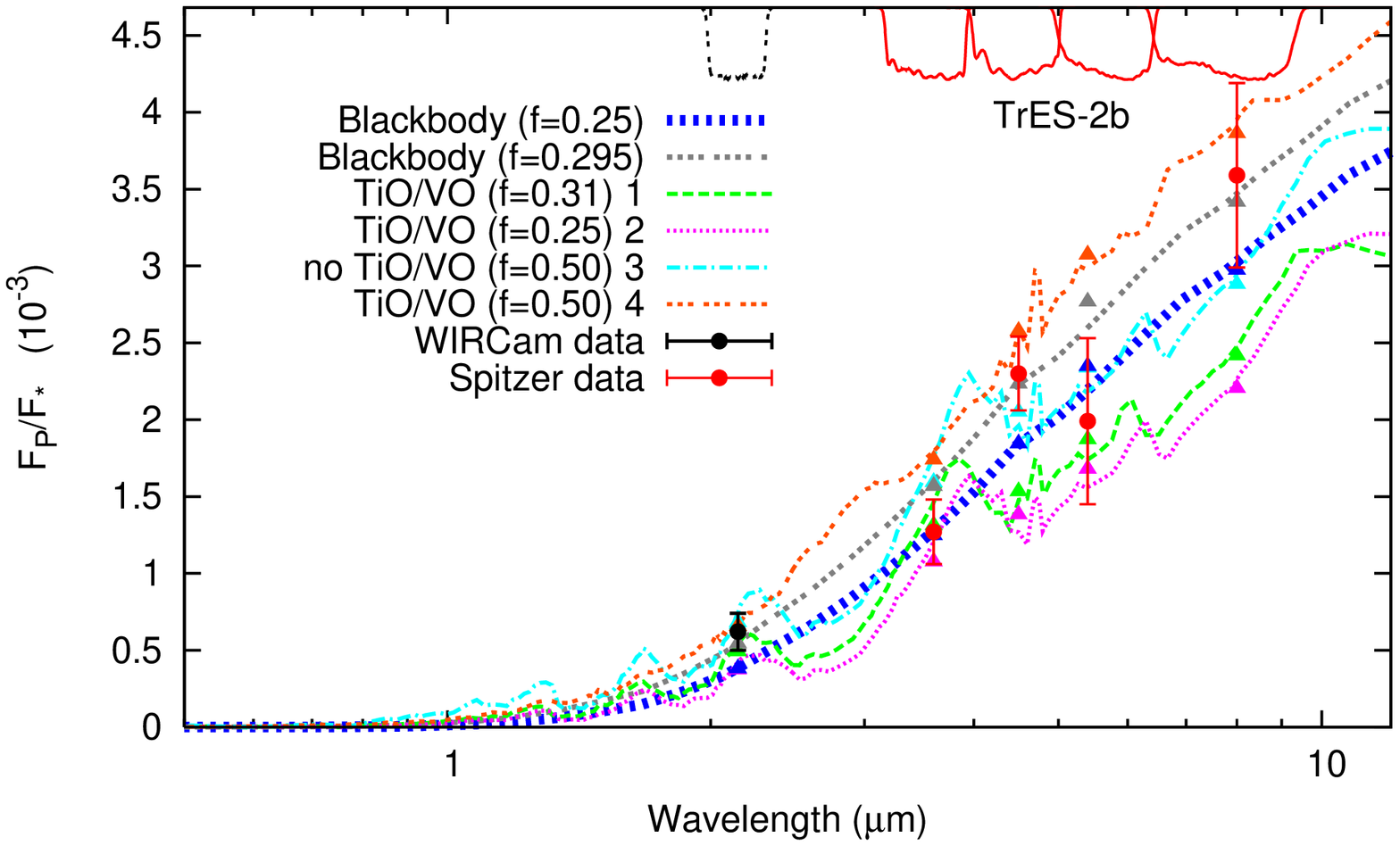}
\includegraphics[scale=0.465,angle=270]{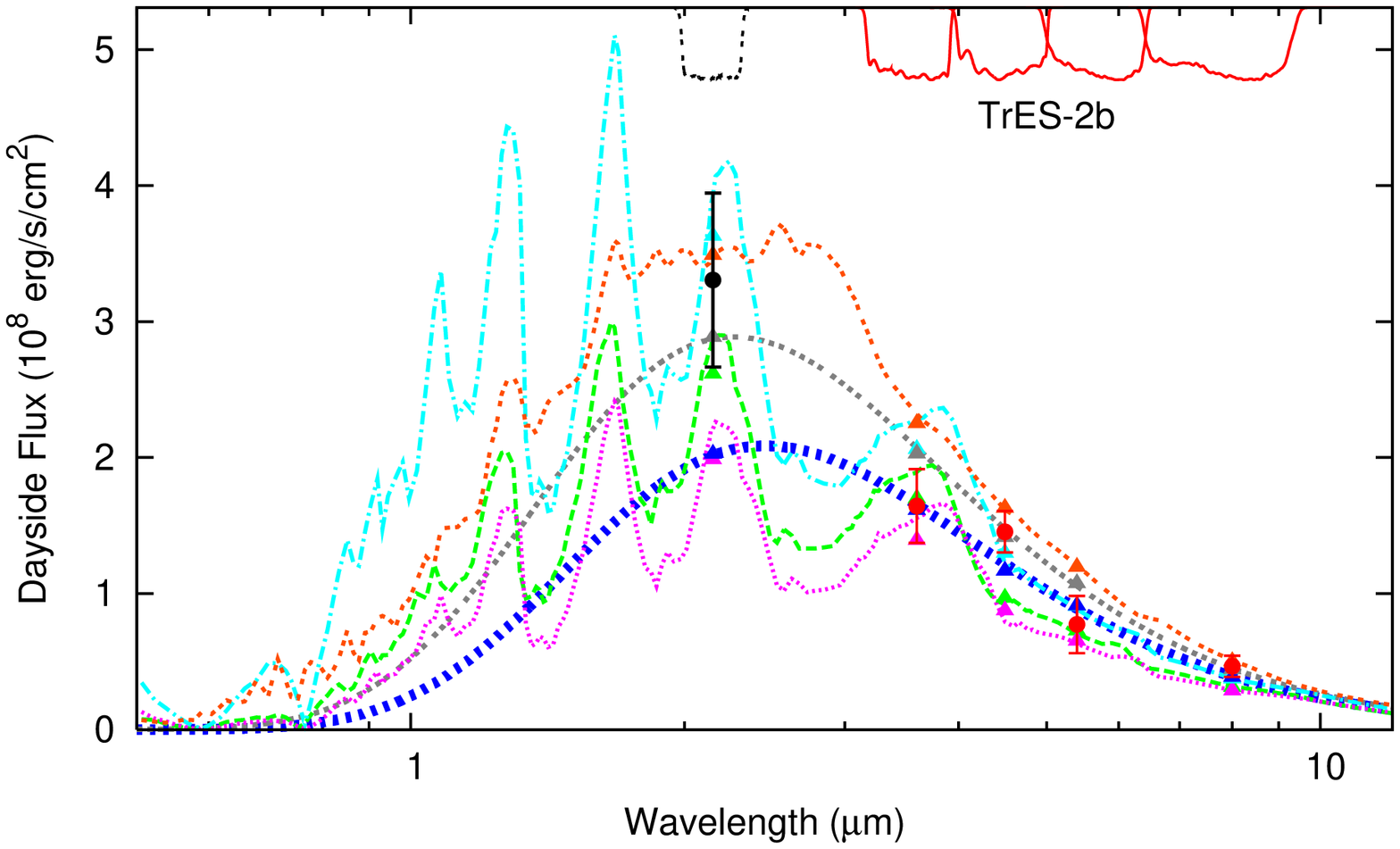}
\caption{	
		Dayside planet-to-star flux ratios (top) and dayside flux at the planet's surface (bottom).
		The Ks-band point (black point; $\sim$2.15 $\mu m$) is our own,
		while the Spitzer/IRAC red points are from \citet{ODonovan09}. 
		Blackbody curves for isotropic reradiation ($f$=$\frac{1}{4}$; $T_{eq}$$\sim$1496 $K$; blue dashed-line)
		and for our best-fit reradiation factor ($f$=0.346; $T_{eq}$$\sim$1622 $K$; grey dotted-line) are 
		also plotted.
		We also plot one-dimensional, radiative transfer spectral models \citep{Fortney06,Fortney08}
		for various reradiation factors and with and without TiO/VO.
		The models with TiO/VO include $f$=$\frac{1}{4}$ (purple dotted line), $f$=0.31 (green dashed line), 
		and $f$ =$\frac{1}{2}$ (orange dotted line); only the last of the models has a temperature inversion.
		The model without TiO/VO features emission from the dayside only ($f$=$\frac{1}{2}$; cyan dot-dashed line).
		The models on the top panel are divided by a stellar atmosphere model \citep{Hauschildt99} of TrES-2 
		using the parameters from \citet{Torres08} ($M_{*}$=0.98 $M_{\odot}$, $R_{*}$=1.00 $R_{\odot}$, $T_{eff}$=5850 $K$, and log $g$= 4.43).
		We plot the Ks-band WIRCam transmission curve (dotted black lines) and Spitzer/IRAC curves
		(solid red lines) inverted at arbitrary scale at the top of both panels.
	}
\label{FigModel}
\end{figure}

We compare the depth of our Ks-band eclipse and the 
Spitzer/IRAC eclipses \citep{ODonovan09} to a series of planetary atmosphere models in Figure \ref{FigModel}.
This comparison is made quantitatively as well as qualitatively by integrating the models over the WIRCam Ks band-pass
as well as the Spitzer/IRAC channels, and calculating the $\chi^{2}$ of the thermal emission data compared to the models.
We first plot blackbody models with an isotropic reradiation factor ($f$=$\frac{1}{4}$; blue dotted-line)
and that of our best-fit value ($f$=0.346; grey dotted-line) these models have dayside temperatures of $T_{day}$$\sim$1496$K$ and 
$T_{day}$$\sim$1622$K$, respectively.
Both blackbody models provide reasonable fits to the data, although the latter model ($f$=0.346; $\chi^{2}$=\BlackbodyTwoChi)
provides a definitively
better fit than the former isothermal model ($f$=$\frac{1}{4}$; $\chi^2$=\BlackbodyOneChi) as it better
predicts our Ks-band emission and the Spitzer/IRAC 8.0 $\mu m$ emission.
This suggests that overall TrES-2b has a near-isothermal dayside temperature-pressure profile
and is well-fit by a blackbody. 

We thus also compare the data to a number of one-dimensional, radiative transfer, spectral models 
\citep{Fortney06,Fortney08} with different reradiation factors that specifically include or exclude gaseous TiO and VO
into the chemical equilibrium and opacity calculations.
In these models when TiO and VO are present they act as absorbers at high altitude and lead to a
hot stratosphere and a temperature inversion \citep{Hubeny03}. However, if the temperature
becomes too cool (TiO and VO start to condense at 1670 $K$ at 1 mbar [\citealt{Fortney08}]),
TiO and VO condense out and
the models with and without TiO/VO are very similar. In the case of TrES-2b, for all the models we calculate,
except our model that features dayside
emission only ($f$=$\frac{1}{2}$), they do not harbour temperature inversions because
the atmospheres are slightly too cool and 
TiO/VO has condensed out of their stratospheres.
We plot models with TiO/VO and reradiation factors of $f$=$\frac{1}{4}$ (purple dotted line), $f$=0.31 (green dashed line), and
$f$=$\frac{1}{2}$ (orange dotted line), and without TiO/VO with a reradiation factor of $f$=$\frac{1}{2}$
(cyan dot-dashed line).
\citet{ODonovan09} argued that TrES-2b experienced a temperature inversion due to the high
4.5 $\mu m$ emission compared to the low 3.6 $\mu m$ emission, which was predicted
to be a sign of water and CO in emission, rather than absorption, in TrES-2b's presumably inverted
atmosphere. We also find that our models without 
a temperature inversion have difficultly matching the Spitzer/IRAC 5.6 and 8.0 $\mu m$ thermal emission
($\chi^{2}$=\FortneyTwoChi \ for $f$=$\frac{1}{4}$ with TiO/VO, $\chi^{2}$=\FortneyOneChi \
for $f$=0.31 with TiO/VO, and $\chi^{2}$=\FortneyThreeChi \ for $f$=$\frac{1}{2}$ without TiO/VO).
If the temperature inversion is due to TiO/VO, by the time the atmosphere becomes hot enough that TiO/VO remains in gaseous form, the thermal 
emission is too bright to fit the 3.6, and 5.8 $\mu m$ thermal emission ($\chi^{2}$=\FortneyFourChi \
for $f$=$\frac{1}{2}$ with TiO/VO).

The combination of our blackbody and radiative transfer models with our own eclipse depth
and those from the Spitzer/IRAC instrument \citep{ODonovan09}
thus suggest that the atmosphere of TrES-2b likely features modest redistribution of heat from the day to the
nightside. It is unclear whether the atmosphere of TrES-2b requires a temperature inversion. 
A simple blackbody model ($f$=0.346 and $T_{eq}$$\sim$1622 K)
provides an exemplary fit to the data; this may indicate that TrES-2b has a 
fairly isothermal dayside temperature structure, perhaps similar to HAT-P-1b \citep{Todorov10}.
An important caveat, on the above result is that our $f$=$\frac{1}{2}$ model without TiO/VO ($\chi^{2}$=\FortneyThreeChi)
and thus without a temperature inversion
returns nearly as good of fit as our best-fit blackbody model ($f$=0.346; $\chi^{2}$=\BlackbodyTwoChi);
thermal emission measurements at other wavelengths,
and repeated measurements at the above wavelengths, are
thus necessary to differentiate
a blackbody-like spectrum, from significant departures from blackbody-like behaviour, and to confirm that TrES-2b efficiently redistributes
heat to the nightside of the planet.  Specifically, the variations between the models displayed in Figure \ref{FigModel}
are largest in the near-infrared J \& H-bands and thus further near-infrared constraints -- if they are able to
achieve sufficient accuracy to measure the small thermal emission signal of TrES-2b in the near-infrared --
should prove eminently useful to constrain the atmospheric characteristics of this planet.

If the excess emission at 4.5 $\mu m$ is due to water emission, rather than absorption,  due to a temperature inversion
in the atmosphere of TrES-2b then the inversion is unlikely to be due to TiO/VO.
This is because the atmosphere of TrES-2b appears too cool to allow TiO/VO
to remain in gaseous form in its upper atmosphere. 
If there is a temperature inversion then
the high altitude optical absorber is likely to be due to another chemical species than TiO/VO.
For instance, \citet{Zahnle09} have investigated
the photochemistry of sulphur-bearing species as another alternative. 

TrES-2b is a promising target for the characterization of its thermal emission across a wide wavelength range.
In addition to orbiting a relatively bright
star, and having a favourable planet-to-star radius ratio, TrES-2 lies within the Kepler field. 
The combination of secondary eclipse measurements already published using Spitzer/IRAC, upcoming 
measurements with Kepler ($\sim$430 - 900 $nm$; \citealt{Borucki08}), and J, H and K-band near-infrared measurements
that could be obtained from the ground, will allow us to fully constrain TrES-2b's energy budget.
At the shorter end of this wavelength range it should also be possible to constrain
the combination of reflected light and thermal emission. Our results predict that even if the geometric albedo of TrES-2b
is as low as 5\% in the Kepler bandpass, if Kepler is able to detect the secondary eclipse of this planet then
it will be detecting a significant
fraction of reflected light in addition to thermal emission. This will largely 
break the degeneracy on the Bond albedo and the reradiation factor for this planet, facilitating a more complete
understanding of its energy budget.

\acknowledgements

The Natural Sciences and Engineering Research Council of Canada supports the research of B.C. and R.J.
The authors would like to thank Marten van Kerkwijk for helping to optimize these observations, and Norman Murray for useful discussions.
The authors especially appreciate the hard-work and diligence of the CFHT staff
in helping us pioneer this ``stare'' method on WIRCam. We thank the anonymous referee for a thorough review.

\end{document}